\documentclass[preprint]{aastex}

\slugcomment{Submitted to ApJ Letters}

\shorttitle{A Li-Rich Giant in NGC 6819}
\shortauthors{Anthony-Twarog, Deliyannis, Rich, Twarog}
\begin{document}

\title{A Lithium-Rich Red Giant Below the Clump in the Kepler Cluster, NGC 6819}
\author{Barbara J. Anthony-Twarog}
\affil{Department of Physics and Astronomy, University of Kansas, Lawrence, KS 66045-7582, USA}
\email{bjat@ku.edu}
\author{Constantine P. Deliyannis}
\affil{Department of Astronomy, Indiana University, Bloomington, IN 47405-7105, USA}
\email{con@astro.indiana.edu}
\author{Evan Rich}
\affil{Department of Physics and Astronomy, University of Kansas, Lawrence, KS 66045-7582, USA}
\email{evan66210@gmail.com}
\and
\author{Bruce A. Twarog}
\affil{Department of Physics and Astronomy, University of Kansas, Lawrence, KS 66045-7582, USA}
\email{btwarog@ku.edu}

\begin{abstract}
WIYN/HYDRA spectra in the Li 6708 \AA\ region have been obtained for 332 probable members of the old open cluster, NGC 6819. Preliminary analysis shows a pattern of Li depletion from the top of the turnoff to the base of the giant branch. Starting one magnitude below the level of the clump, all brighter giants have A(Li) below 1.0, with most having upper limits below 0.5. Star W007017, located {\it below} the first-ascent red giant bump is Li-rich with A(Li) = 2.3. As a highly probable single-star astrometric and radial-velocity cluster member, its discrepant asteroseismic membership could be a by-product of the processes that triggered Li-enhancement. Its color-magnitude diagram location is consistent with only one proposed enhanced mixing process among first-ascent red giants.  
\end{abstract}



\section{Introduction}
Elemental abundances in stars supply insight into astrophysical processes governing Galactic chemical evolution and the structural evolution of stars. Few elements have garnered as much attention for their impact on cosmology and stellar evolution as Lithium. Created during Big Bang nucleosynthesis, the primordial Li abundance acts as a sensitive constraint for models of the early Universe. Easily destroyed in stellar interiors at T $\geq 2.5 \times 10^6$ K, its variation with evolutionary phase is a diagnostic of the temporal significance of mixing and/or diffusion in stellar atmospheres. Not surprisingly, both areas suffer from discrepancies between standard theory and observations. The WMAP estimate \citep{LA11} for A(Li) (= 12 + log(N$_{Li}$/N$_H$)) is three times higher than the observed limit in halo dwarfs \citep{SP12} while disk field and open cluster stars show Li depletion where none should exist, e.g. the Li-dip among F dwarfs \citep{BT86}, or depletion rates that far exceed predicted values, e.g. among cooler dwarfs like the Sun \citep{XD09}.

Assuming the WMAP-inferred value for the Big Bang Li abundance, a 4-fold increase in Li between Big Bang nucleosynthesis and solar system formation requires major Li sources on Galactic scales. Cosmic ray spallation remains the only viable source for production of $^6$Li, but light element isotopic abundances require a stellar origin for at least 70$\%$ of the $^7$Li associated with the solar system \citep[][and references therein]{PR12}. The need to create Li in a non-destructive but observationally accessible environment has led to an emphasis on abundance anomalies among evolved stars, i.e. a search for the atmospheric signatures of the Cameron-Fowler (CF) mechanism \citep{CF71} generating Li in low-mass red giants via cool-bottom burning \citep{SB99} and in intermediate and higher mass asymptotic giant branch stars via hot-bottom burning \citep[][and references therein]{KA11}. To date, no consensus has emerged on the efficacy of either process. Searches for super-Li-rich giants, stars with A(Li) above the current ISM value of $3.3$, and Li-rich giants, stars with A(Li) above $1.5$, the standard stellar evolution limit for lower mass stars ascending the giant branch after the first dredge-up phase \citep{KU11, UT12}, typically have an incidence rate below 1\% \citep{SM10}. In recent years, giants with anomalous Li abundances have been identified in the Galactic Bulge \citep{LB12}, dwarf spheroidal galaxies \citep{KI12}, the thick disk \citep{MO11}, and the field star population \citep{KU11}. While \citet{CB00} and \citet{KU11} have claimed that Li-rich giants concentrate in a narrow luminosity range associated with the red giant bump and/or the red giant clump, other investigations imply that Li-rich giants can be found almost anywhere along the red giant branch. This ambiguity makes assigning responsibility for Li enhancement difficult and is partially due to the range in mass, chemical composition, and evolutionary phase found among the mixed population of stars included in most surveys. Ideally, one would prefer only stars with a range in mass at a given age and composition or stars of a given mass and composition over a range in age, making star clusters ideal contexts for these studies.

With the goal of probing stellar structure and evolution among low mass stars via atmospheric Li, the authors have undertaken an extensive program to survey spectroscopically members of a key set of open clusters from the tip of the giant branch to as far down the main sequence as technology allows. First results have been published for the clusters NGC 3680 (age = 1.75 Gyr) \citep{AT09} and NGC 6253 (3.0 Gyr) \citep{AT10, CU12}. Among the clusters currently under analysis are NGC 7789 (1.5 Gyr) and NGC 6819 (2.3 Gyr), with over 300 stars observed in each. NGC 6819 has the distinction of lying within the Kepler field and remains the focus of asteroseismic studies reaching down the giant branch \citep{ST11}. The purpose of this Letter is to report the discovery of a unique Li-rich giant in NGC 6819. A detailed investigation of the theoretical implications of the Li pattern in NGC 6819, as well as the Li-rich giant, are beyond the scope of this Letter and will be discussed in a future paper \citep{AT13}. We will, however, provide some context to clarify the unique nature of this highly probable cluster member.

\section{Observations and Reduction}
Spectra of 332 probable members of the open cluster NGC 6819 were obtained between 2010 and 2012 using the WIYN 3.5m telecope at Kitt Peak National Observatory equipped with the HYDRA multi-object fiber-based spectrograph. Six configurations were used to extend the sample from $V$ = 11 to 16.5. Complete details of the observation and reduction procedures will be provided as part of a comprehensive discussion of the spectra and Li abundances for all 332 stars in \citet{AT13}. The reader is referred to \citet{CU12} for insight into the calibration and reduction steps used within the project.

The cluster spectra cover a wavelength range from 6550 to 6840 \AA\, with a resolution of 0.5 \AA\ = 2.7 pixels. Due to the plethora of lines for cool giants at the metallicity of NGC 6819, we have estimated the average S/N per pixel in this wavelength region based on the accumulated photon statistics. While these probably optimistic S/N ratios per pixel for the entire sample range from below 70 to several hundred, the evolved stars discussed in the current Letter have S/N per pixel above 100, with a majority above 200.

\section{The Color-Magnitude Diagram and Li Evolution}
Sample selection was based on the comprehensive photometric and radial-velocity study of the cluster by \citet{HO09}(H09). All stars chosen were classed by H09 as radial-velocity members with a probability above 50$\%$. The color-magnitude diagram (CMD) using photometry from Table 2 of H09 for our 332 stars is shown in Fig. 1. The magenta starred point is the proposed Li-rich giant. Superposed is a $Y^2$ \citep{DE04} isochrone with an age of 2.25 Gyr and [Fe/H] = +0.09, adjusted for $E(B-V)$ = 0.14 and $(m-M)$ = 12.5. The $Y^2$ isochrones have been adopted for consistency with previous cluster studies \citep{AT09, AT10, CA11}. 

We have adopted the reddening value and metallicity derived by \citet{BR01} from high-dispersion spectroscopy of four clump giants. $E(B-V) = 0.14$ falls within the range $E(B-V)$ of 0.12 to 0.16 found in published studies of the cluster using methods such as differential cluster CMD matches to isochrones and M67 \citep{RV98}, Washington photoelectric photometry \citep{CA86}, and UBV photoelectric and photographic data \citep{BU71}. \citet{KA01} adopt $E(B-V)$ = 0.10. The adopted metallicity ([Fe/H]$ = 0.09$) from \citet{BR01} is consistent with the adopted reddening; moderate-dispersion spectroscopy of seven giants by \citet{FR02} using $E(B-V)$ = 0.16 gives [Fe/H] = 0.04 if their scale is adjusted so that M67 has [Fe/H] = 0.0 \citep{TW97}. Ultimately, our conclusions are remarkably insensitive to the value of the reddening (see Sec. 4). 

With the reddening and metallicity known, $T_{eff}$ was derived for each star using the $B-V$ of Fig. 1 and the color-temperature-metallicity relations for cool giants from \citet{AA96}. Surface gravities were obtained by mapping $V$ to 
{\it log g} values established by the isochrone comparison (Fig. 1). For 21 stars with mass and radius information from the Kepler analyses by \citet{ST11}, we compared the isochrone-derived surface gravities with those from asteroseismology and accordingly incremented the isochrone-implied values by 0.08. 

Equivalent (EW) measures were made by multiple analysts using SPLOT to measure lines within 10 \AA\ of the Li 6708 \AA\ feature, then combined to a common measurement system. After subtracting the expected contribution to the 6708 \AA\ feature that is due to Fe I at 6707.45 \AA\ \citep{SD04}, we estimate the Li abundance by interpolating within a grid of temperature and Li abundances from \citet{ST03}.

While this approach works well for stars at the turnoff, it is less satisfactory for evolved stars whose temperatures are near the low temperature limit of the grid's applicability. Also, the line-dense spectra of the giants complicates placement of the stellar continuum near the typically weak Li line, so we utilized spectrum synthesis techniques.  An appropriate model atmosphere for each star is specified by its color-based $T_{eff}$, its $V$-dependent $log$ $g$ and a microturbulent velocity estimated from surface gravity as follows:  $v_t = 2.0 -0.2$ $log$ $g$ following \citet{CA04} and \citet{MB08}. Using MOOG, comparison of the synthesized spectrum, for which the Li abundance may be varied interactively, to the observed spectrum leads to an optimal A(Li). We stress that for the majority of giants, the 6708 \AA\ feature is dominated by the nearby Fe I line, so only an upper limit is possible for Li. The upper limits on A(Li) are conservative in that the limit would need to be reduced by at least 0.3 dex for the synthetic spectrum to approach the observed line.

Fig. 2a shows the CMD for all evolved stars redder than $B-V$ = 0.70. The caption summarizes symbol keys that provide information on binarity as well as highlighting asteroseismic non-members, subgiants and the proposed Li-rich giant.   Fig. 2b shows the same stars with the measured A(Li) as the primary axis. The mean value for the stars at the turnoff above the Li-dip is marked by the vertical arrow at A(Li) = 3.3. Fig. 3 illustrates the spectra for W007017, with an upper limit of Vsini = 8.3 $\pm$ 0.3 km/sec from IRAF's {\it fxcor}, and a cluster member with nearly identical color and temperature, W005011.

\section{Discussion and Conclusions}
Ignoring binaries and/or potential non-members, the evolutionary pattern in Fig. 2 resembles that outlined in the younger clusters NGC 752, NGC 3680, and IC 4651 \citep{PA04, AT09}. Unevolved stars at the turnoff on the hot side of the Li-dip region have the solar system value of A(Li) = 3.3 and are thus inferred to be depleted minimally, if at all. As stars evolve across the subgiant branch, the SCZ deepens drastically. A(Li) is depleted by standard dilution \citep{DE90} and/or non-standard mixing to below 1.5 when the SCZ reaches its maximum depth. For stars with $V\leq 14$, there is a precipitous drop in A(Li); typical upper limits of A(Li) = 0.5 or less are found. As expected due to the $T_{eff}$ sensitivity of the Li feature (and higher S/N), the upper limits implied by a given line strength decline as stars become redder. Only 6 stars with $V \leq 14$, including two known binaries, have measurable A(Li). All but one have A(Li) below 1.0. The Li-rich giant, W007017 (WOCS ID from H09), has A(Li) = 2.3 and is located below the red giant clump and the red giant bump by $\sim$0.5 mag. From standard stellar evolution, this implies that the star must be a first-ascent red giant with an inert He-core surrounded by a H-burning shell. The latter comparison is critical because the red giant bump marks the phase where the H-burning shell crosses the chemical discontinuity left behind by the bottom of the convective envelope at the end of the first dredge-up. Above this luminosity level, lower mass stars are believed to undergo extra mixing in their convectively stable radiative zones between the H-burning shell and the bottom of the convective envelope \citep{RD95}, leading to enhanced atmospheric Li as a by-product of the CF mechanism. The CMD position of this star is incompatible with all current models for Li enhancement in low mass giants \citep[][and references therein]{CL10} except one. Before discussing the exception, a few issues relevant to the paradoxical CMD location of the Li-rich giant should be addressed.  

a) Is W007017 a member of NGC 6819? It is a proper motion (90$\%$) \citep{SA72} member, a single-star radial-velocity (95$\%$) \citep{HO09}member, and a photometric member (located at the red edge of the RGB).  The first two together imply near-certainty of cluster membership; furthermore, it is even more improbable that a non-member would meet all {\it three} of these criteria by coincidence. And yet, \citet{ST11} have classified it as a seismic non-member. Cluster membership based upon seismic oscillations is a measure of the degree of similarity in the oscillation pattern among stars at the same position/evolutionary phase in the CMD. Four of the six seismic non-members in NGC 6819 are also photometric non-members. The two photometric members, W007021 and W007017, lie along the red edge of the first-ascent giant branch, 0.5 to 0.7 mag below the level of the clump. Of the six seismic non-members, W007021 exhibits the greatest deviation from the relations defined by the cluster members while W007017 shows the smallest. Seismic oscillations have the potential to identify as non-members any field interlopers whose kinematics mimic that of the cluster, but also {\it any exotic stars, such as remnants of strong dynamic interactions} \citep{ST11} because standard evolution is assumed for cluster stars when estimating seismic parameters. While it isn't possible to entirely rule out the possibility that W007017 is be a Li-rich field star mimicking the cluster space motion and photometric sequence, it seems far more probable that its seismic deviation from the cluster relations reflects the structural mechanism that led to its anomalously high atmospheric Li abundance.

b) How reliable are the cluster/stellar parameters? A(Li) depends upon the derived $T_{eff}$, $log$ $g$, and $v_t$, in turn dependent upon the adopted [Fe/H], reddening, distance modulus, and age. To test the sensitivity of our conclusions to these parameters, A(Li) was rederived using synthetic spectra with $T_{eff}$ altered by $\pm$100 K, $log$ $g$ by $\pm$0.1, and [Fe/H] by $\pm$0.1 dex. While systematic shifts occurred in the absolute scale of Fig. 2b, the relative A(Li) ranking of the stars remained the same within the measurement errors. As an example, lowering $T_{eff}$ by 100 K on average dropped A(Li) by 0.1 dex, with a smaller shift of 0.05 dex near the turnoff, increasing to 0.15 dex among the cooler giants; W007017 still remained anomalously rich (A(Li) = 2.2) compared to the evolved stars brighter than $V$ = 14.0.   
A related question is tied to the discrepancy between the position of W007017 in the CMD and the red giant bump. As the cluster ages and lower mass stars evolve up the giant branch, the luminosity of the bump location drops and the luminosity range of the bump grows larger \citep{DE04}. Would increasing the age of the cluster place the Li-rich giant above the bump? As implied by Fig. 2a, increasing the age of the cluster from 2.25 Gyr to 2.5 Gyr would require a reddening estimate reduced to $E(B-V)$ = 0.10 to match the turnoff color. In turn, lowering the reddening would lower the apparent modulus to 12.3, the same combination of values derived by \citet{KA01} using fits to the Hyades cluster. So, while the $M_V$ limit for the bump would be about 0.2 mag fainter for the older giant branch, the $V$ limit would remain near 13.0, unchanged from the current estimate.

c) Is W007017 a composite system? Of the 6 stars with measurable Li, 5 lie within 1 mag of each other near the clump and, of these, two are known binaries. While the measurements to date number 5, the radial-velocity data of H09 and the current study show no evidence for radial-velocity variations or unusual line broadening, so the data are consistent with a single star.
It should be noted that two fainter, bluer stars lie within 5\arcsec\ of W007017. The brighter star at $V$ = 16.69 and $B-V$ = 0.81 is 3.3\arcsec\ away while the fainter star at $V$ = 17.18 and $B-V$ = 0.70 is 4.9\arcsec\ removed, while W007017 has $V$ = 13.57 and $B-V$ = 1.24. Since the CCD photometry used in the determination of $V$ and $B-V$ for W007017 is based upon point-spread-function fitting techniques, it is unlikely that the colors for W007017 are seriously impacted by the presence of this pair. If, however, all three stars were erroneously measured as one system, the composite would be brighter by 0.1 mag and bluer by 0.05. More realistic mixtures of the contaminating stars scaled by the square of their distance from W007017 lead to upper limits on the impact of these two stars of -0.05 mag in $V$ and -0.03 mag in $B-V$. Therefore, in the limit, W007017 could have true colors closer to $V$ = 13.62 and $B-V$ = 1.27. A shift in $B-V$ by 0.03 alters the $T_{eff}$ of W007017 by 66 K and lowers the derived A(Li) by less than 0.1 dex. For the seismic analysis, the contaminating pair would increase the noise of the signal but not alter the oscillation frequencies \citep{ST11}.

Although the evidence for cluster membership is compelling, W007017 is in any case a member of the rare class of red giants with anomalously high atmospheric Li. If a member of NGC 6819, its position in the CMD places it below both the red giant clump and red giant bump for first-ascent red giants, leaving the obvious question of the origin of the enhanced Li. If there is no interacting companion, an extremely salient possibility emerges from the recent work of \citet{DE12}. In attempting to explain the apparent distribution of red giants below the bump as compiled by \citet{KU11}, \citet{DE12} found that single star models undergoing fast internal rotation could experience extra turbulent mixing, resulting in enhanced surface Li at the level found in W007017.  There could also be extra heat transport, affecting the structure in the radiative zone above the H-burning shell (by increasing dlnT/dlnP) as the star evolves brighter than the red giant bump luminosity, causing the star to zigzag toward lower luminosity along a cooler evolutionary track before reascending the giant branch. Thus, Li-rich giants on the first-ascent of the giant branch should be positioned redward of the standard giant branch, fainter than the red giant bump, and should have an atmospheric structure that differs significantly from standard giants ascending to the red giant bump for the first time. Such stars could be relatively rare, since only the most rapidly rotating stars would become Li-rich and execute zigzags.  Whether this scenario applies to W007017 remains to be seen, but the unique properties of this star indicate that additional scrutiny, particularly searching for
spectroscopic signatures of exceptional mixing, such as, for example, the 
$^{12}{\rm C/^{13}C}$ ratio, could be beneficial to our understanding of this rare class of objects. 

\acknowledgments
It is a pleasure to acknowledge the line measurements contributed by undergraduates Jeremy Ims and Daniel Webb of KU which aided the analysis of the spectra, as well as the observational support provided by then grad students Ryan Maderak and Jeff Cummings of IU in collecting the spectra. NSF support for this project was provided to ER through NSF grant AST-0850564 via the CSUURE REU program supervised by Eric Sandquist, to BJAT and BAT through NSF grant AST-1211621, and to CPD through NSF grant AST-1211699. Useful feedback on the interpretation of the asteroseismology was supplied by Ron Gilliland and Dennis Stello. Extensive use was made of the WEBDA database maintained by E. Paunzen at the University of Vienna, Austria (http://www.univie.ac.at/webda) and of the MOOG suite of spectroscopic analysis software.

\clearpage
\begin{figure}
\includegraphics[angle=270,scale=0.6]{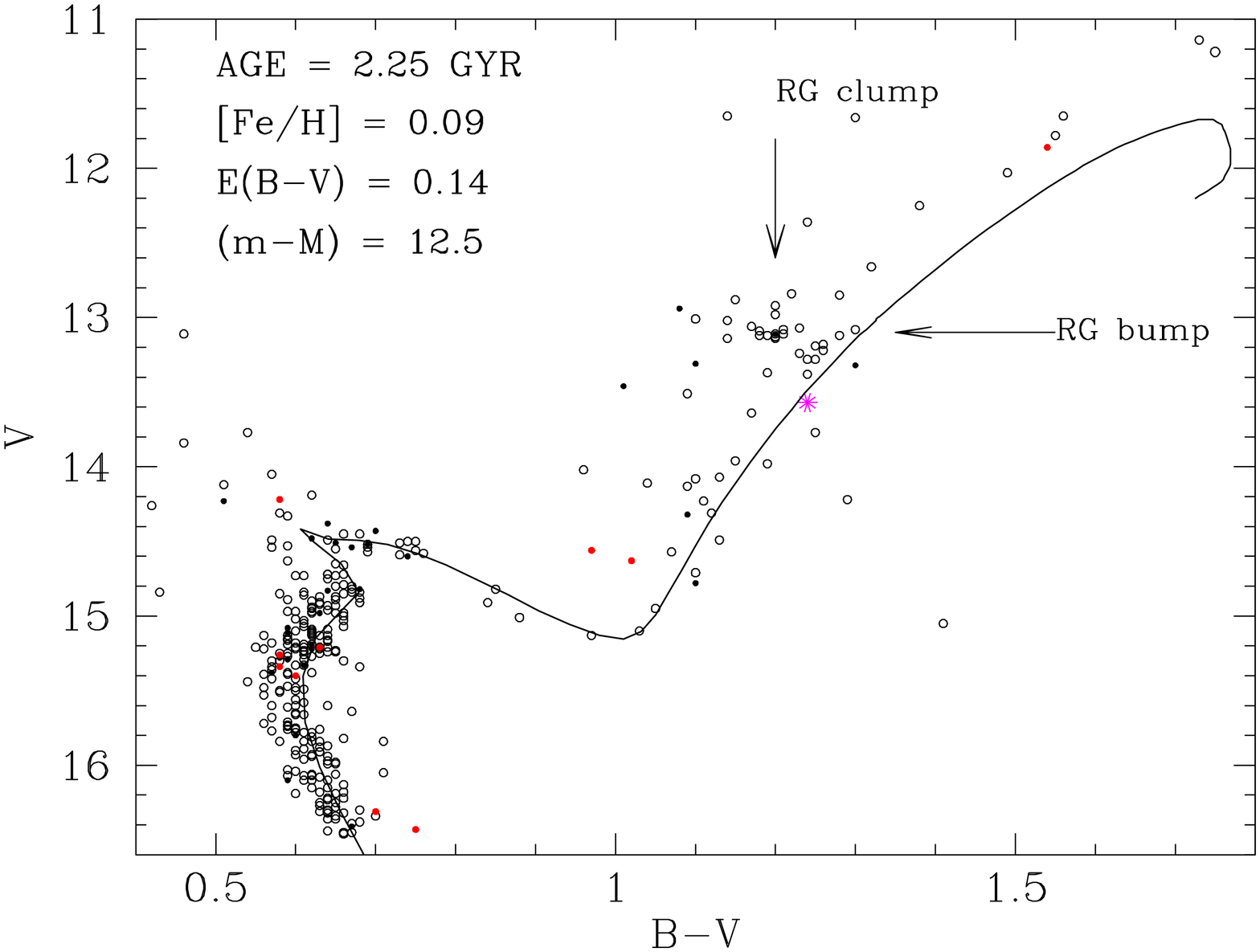}
\caption{CMD of stars observed spectroscopically within NGC 6819. Open circles are single-star members, as verified by the radial-velocity study by \citet{HO09}.  Filled symbols denote probable binaries, as noted by \citet{HO09}, black and the present study, red. The starred magenta point marks the position of the proposed Li-rich giant. Superposed is a $Y^2$ \citep{DE04} isochrone with an age of 2.25 Gyr and [Fe/H] = +0.09, adjusted for $E(B-V)$ = 0.14 and $(m-M)$ = 12.5.}
\end{figure}
\clearpage
\begin{figure}
\includegraphics[angle=270,scale=0.6]{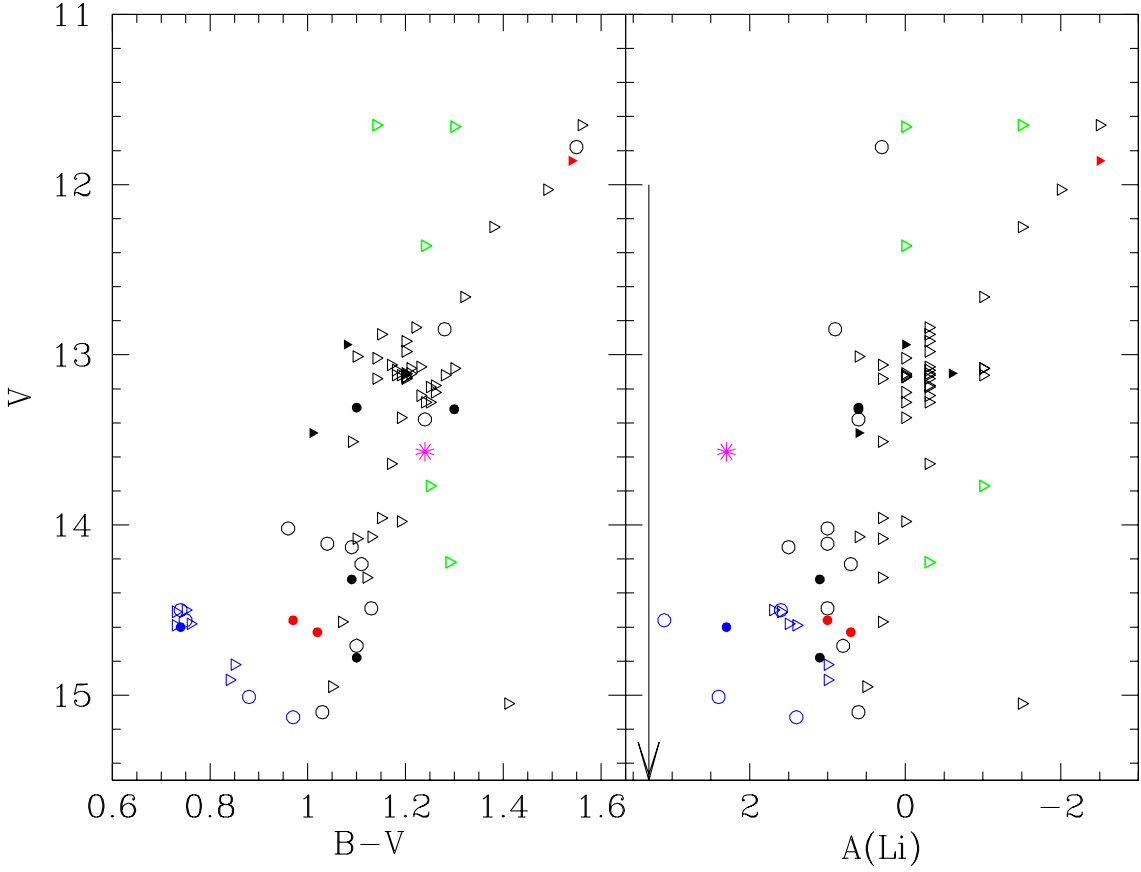}
\caption{ Left panel: CMD of the evolved stars ($(B-V) \geq 0.70$) with derived A(Li). Open circles are single-star members with measurable Li; open triangles are stars with upper limits for A(Li). Filled symbols are probable binaries fro H09 (black) and the current study (red). Green symbols are asteroseismic non-members from \citet{ST11}. Blue symbols delineate subgiant stars. The magenta starred point is the Li-rich giant. Right panel: A(Li) for stars along the CMD. Symbols have the same meaning as in the left panel.}
\end{figure}
\begin{figure}
\includegraphics[angle=270,scale=0.6]{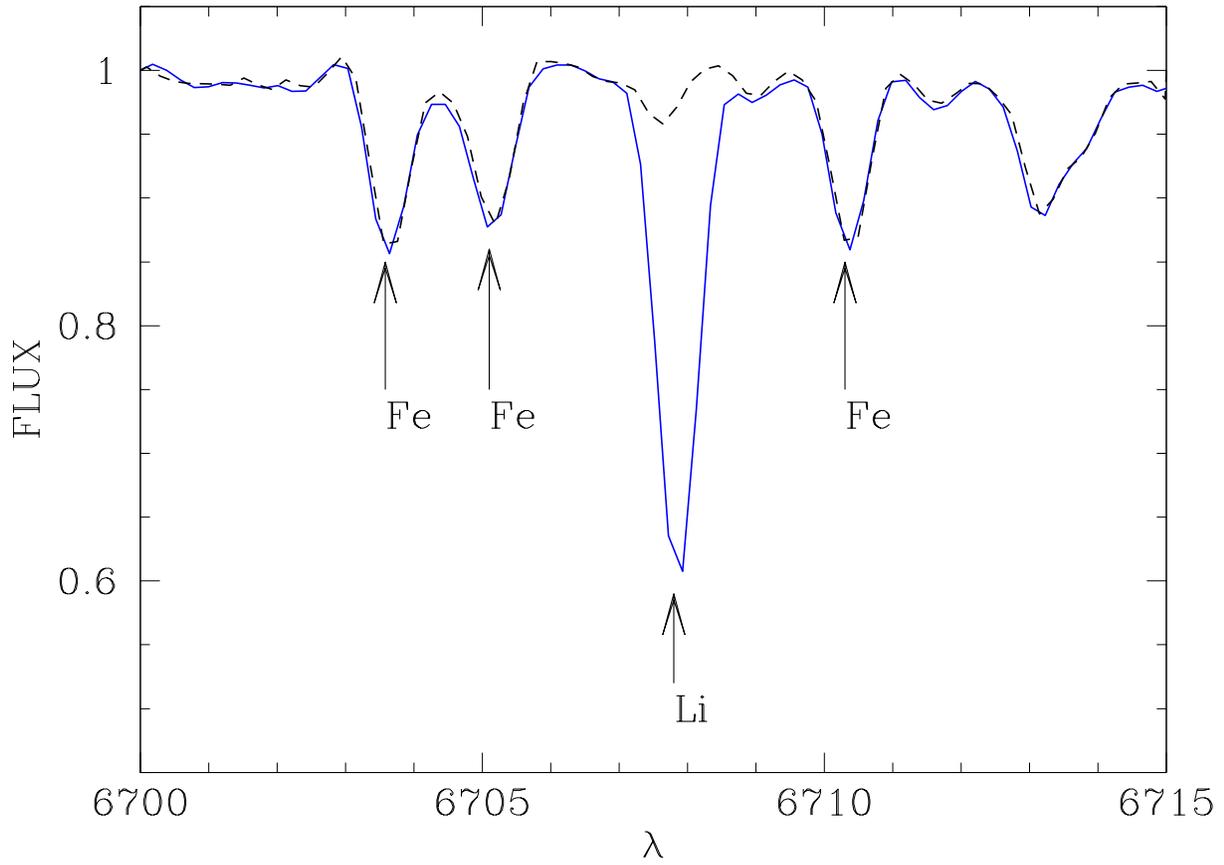}
\caption{Spectra near the 6707.8 \AA\ line for red giants W007017 (blue) and W005011.  Note the Fe I line at 6707.45 \AA\ in W005011.}
\end{figure}
\end{document}